\documentclass[prb,twocolumn,showpacs,floatfix,amsmath,amssymb,superscriptaddress]{revtex4-1}
\usepackage{amsfonts}
\usepackage{stmaryrd}
\usepackage{bbm}
\usepackage{mathrsfs}
\usepackage{tipa}
\usepackage{amssymb}
\usepackage{txfonts}
\usepackage{graphicx}
\usepackage{dcolumn}
\usepackage{epstopdf}
\usepackage[colorlinks,linkcolor=blue,urlcolor=blue,citecolor=blue]{hyperref}
\usepackage{multirow}
\usepackage{subfigure}
\usepackage{url}

\begin{document}

%\preprint{APS/123-QED}

\title{Optical spectroscopy and ultrafast pump-probe study of a quasi-one-dimensional charge density wave in CuTe}% Force line breaks with \\

\author{R. S. Li}
\affiliation{International Center for Quantum Materials, School of Physics, Peking University, Beijing 100871, China}
\author{L. Yue}
\affiliation{International Center for Quantum Materials, School of Physics, Peking University, Beijing 100871, China}
\author{Q. Wu}
\affiliation{International Center for Quantum Materials, School of Physics, Peking University, Beijing 100871, China}
\author{S. X. Xu}
\affiliation{International Center for Quantum Materials, School of Physics, Peking University, Beijing 100871, China}
\author{Q. M. Liu}
\affiliation{International Center for Quantum Materials, School of Physics, Peking University, Beijing 100871, China}
\author{Z. X. Wang}
\affiliation{International Center for Quantum Materials, School of Physics, Peking University, Beijing 100871, China}
\author{T. C. Hu}
\affiliation{International Center for Quantum Materials, School of Physics, Peking University, Beijing 100871, China}
\author{X. Y. Zhuo}
\affiliation{International Center for Quantum Materials, School of Physics, Peking University, Beijing 100871, China}
\author{L. Y. Shi}
\affiliation{International Center for Quantum Materials, School of Physics, Peking University, Beijing 100871, China}
\author{S. J. Zhang}
\affiliation{International Center for Quantum Materials, School of Physics, Peking University, Beijing 100871, China}
\author{D. Wu}
\affiliation{Songshan Lake Materials Laboratory, Dongguan, Guangdong 523808, China}
\author{T. Dong}
\affiliation{International Center for Quantum Materials, School of Physics, Peking University, Beijing 100871, China}
\author{N. L. Wang}
\email{nlwang@pku.edu.cn}
\affiliation{International Center for Quantum Materials, School of Physics, Peking University, Beijing 100871, China}
\affiliation{Beijing Academy of Quantum Information Sciences, Beijing 100913, China}

\date{\today}% It is always \today, today,

\begin{abstract}

CuTe is a two-dimensional (2D) layered material, yet forming a quasi-one-dimensional (quasi-1D) charge-density-wave (CDW) along the a-axis in the ab-plane at high transition temperature $T_{CDW}=335$ K. However, the anisotropic properties of CuTe remain to be explored. Here we performed combined transport, polarized infrared reflectivity, and ultrafast pump-probe spectroscopy to investigate the underlying CDW physics of CuTe. Polarized optical measurement clearly revealed that an energy gap gradually forms along the a-axis upon cooling, while optical evidence of gap signature is absent along the b-axis, suggesting pronounced electronic anisotropy in this quasi-2D material. Time-resolved optical reflectivity measurement revealed that the amplitude and relaxation time of photo-excited quasiparticles change dramatically across the CDW phase transition. Taking fast Fourier transformation of the oscillation signals arising from collective excitations, we identify the 1.65-THz mode as the CDW amplitude mode, whose energy softens gradually at elevated temperatures. Consequently, we provide further evidence for the formation of completely anisotropic CDW order in CuTe, which is quite rare in quasi-2D materials.

\end{abstract}

\pacs{Valid PACS appear here}% PACS, the Physics and Astronomy
\maketitle

%\tableofcontents

\section{\label{sec:level1}INTRODUCTION}
Charge density waves (CDWs) have been of great interest in condensed matter physics for several decades. As the name implies, the conducting electrons form a new periodic modulation with a period that is commensurate or incommensurate with the underlying lattice. This phenomenon is usually induced by Fermi surface nesting (FSN), where the Fermi surface tends to nest between nearly parallel parts connected by a vector $\textbf{q}=2\textbf{k}_F$, which causes the electronic susceptibility to diverge at the nesting wave vector~\cite{peierls1955quantum}. Meanwhile, the lattice responds to the electronic condensate through electron-phonon coupling. At $\textbf{q}=2\textbf{k}_F$, acoustic phonon tends to be softened to zero frequency, which further gives rise to the periodic modulation of the lattice structure~\cite{gruner2018density, Ando_2005, PhysRevB.77.165135, PhysRevB.71.085114}.  In the CDW state, a single-particle energy gap will open in the nested region of the Fermi surface, leading to the lowering of the electronic energy of the system. Therefore the experimental characterization of the single-particle energy gap would be a key step in identifying the CDW state, which can be realized by optical conductivity spectroscopy. Besides single-particle excitation, CDWs also have collective excitations consisting of amplitude mode and phase mode. The amplitude mode corresponds to the ionic displacement with a finite gap at $\textbf{q}=0$ limit and behaves as optical phonons~\cite{RevModPhys.46.83, PhysRevB.90.104414}. In this regard, measurement of amplitude mode can also be important evidence of CDW phase transition, which has been verified by both pump-probe and Raman spectroscopy~\cite{PhysRevLett.118.107402, PhysRevB.101.205112,  PhysRevLett.37.1407, PhysRevB.91.144502}.

Generally, CDWs tend to exist in low-dimensional systems. Typical examples include K$_{0.3}$MoO$_3$~\cite{PhysRevB.43.8421}, transition-metal metal dichalcogenides~\cite{PhysRevLett.37.1407, Smith_1985}, rare-earth tritellurides~\cite{PhysRevB.81.073102, PhysRevB.89.075114}, etc. Among all these materials, the underlying CDW mechanism matches very well with the established theory, providing standard systems to study new CDW materials. Recently, the compound CuTe was reported to be a new  CDW material, which has attracted attention for it is a rare 2D layered material hosting quasi-1D CDW order. Its structure is similar to the well-known prototype Fe-based superconductor FeSe. There exists a small difference between in-plane lattice parameters a and b, resulting in an orthorhombic lattice structure. It exhibits period modulation corresponding to Te chains at low temperature, indicating a CDW phase transition~\cite{stolze2013cute}. Angle-resolved photoemission spectroscopy (ARPES) has identified CuTe as a CDW metal with an anisotropic gap below transition temperature 335 K~\cite{zhang2018evidence}. The nested Fermi surface that drives CDW transition is observed to emerge from nearly parallel line Fermi segments, which is connected by CDW wave vector $\textbf{q}_x=0.4 a^*$. Based on the Seebeck coefficient, thermal conductivity, and specific heat measurement, Kou et al. reported that CDW order can have a great influence on the thermal transport properties, due to strong electron-phonon coupling~\cite{kuo2020transport}. Additionally, by application of external pressure, the CDW transition temperature $T_{CDW}$ of CuTe  decreases gradually~\cite{wang2021pressure}. With increasing pressure, superconductivity emerges at 2.4 K after the CDW gets suppressed completely, suggesting that the CDW order competes with superconductivity in CuTe~\cite{wang2021pressure}. Despite these experimental results, the electrical and optical response of CuTe along the b-axis remains unclear. It is highly interesting to investigate how the CDW phase transition affects the anisotropic properties of the 2D layered material, in particular on the transport and optical conductivity properties. 

In this work, we use a combination of transport, optical spectroscopy, and ultrafast pump-probe reflectivity study on the single crystalline CuTe. Optical spectroscopy is a valuable experimental method that probes the bulk properties of materials, including charge dynamics and electronic band structure of a system. On the other hand, ultrafast spectroscopy can probe coherent vibrational dynamics with high resolution to approach a low energy scale. Furthermore, the energy gap arising from CDW order plays an important role in the relaxation process of quasiparticles. The relaxation time and amplitude of quasiparticles show strong enhancement near transition temperature, which could also be probed from the pump-probe measurement. From optical spectroscopy, we find that there appears an energy gap along a-axis which dramatically reduces the carrier density across $T_{CDW}$, while it behaves as a usual metal along the b-axis without showing anomaly below $T_{CDW}$.  The amplitude and the relaxation of photoinduced carriers exhibit a quasidivergence behavior, yielding further evidence for CDW energy gap formation. Moreover, photoinduced reflectivity displays strong oscillation signals on top of the decay process at low temperatures. One of the oscillations has a much higher signal intensity and its energy softens by over 18$\%$ upon heating from 4.6 K to 280 K. The temperature dependence of the oscillation suggests that it arises from the amplitude mode of CDW collective excitation.

\section{\label{sec:level2}EXPERIMENTAL RESULTS}

Single crystals of CuTe were grown by melting stoichiometric amounts of Cu and Te at 500$^{\circ}$C. After 24 h, the melt was cooled to 400 $^{\circ}$C at a rate of 4 $^{\circ}$ C/h. Then the furnace was shut down and cooled to room temperature. The crystal has a flake-like shape with a shining a-b plane as the cleavage plane, consistent with its quasi-2D structure. We can identify crystallographic a-axis perpendicular to the strip-like structure, determined using a Laue X-ray diffractometer.

\begin{figure}[htbp]
	\begin{center}
		\includegraphics[clip, width=0.4\textwidth]{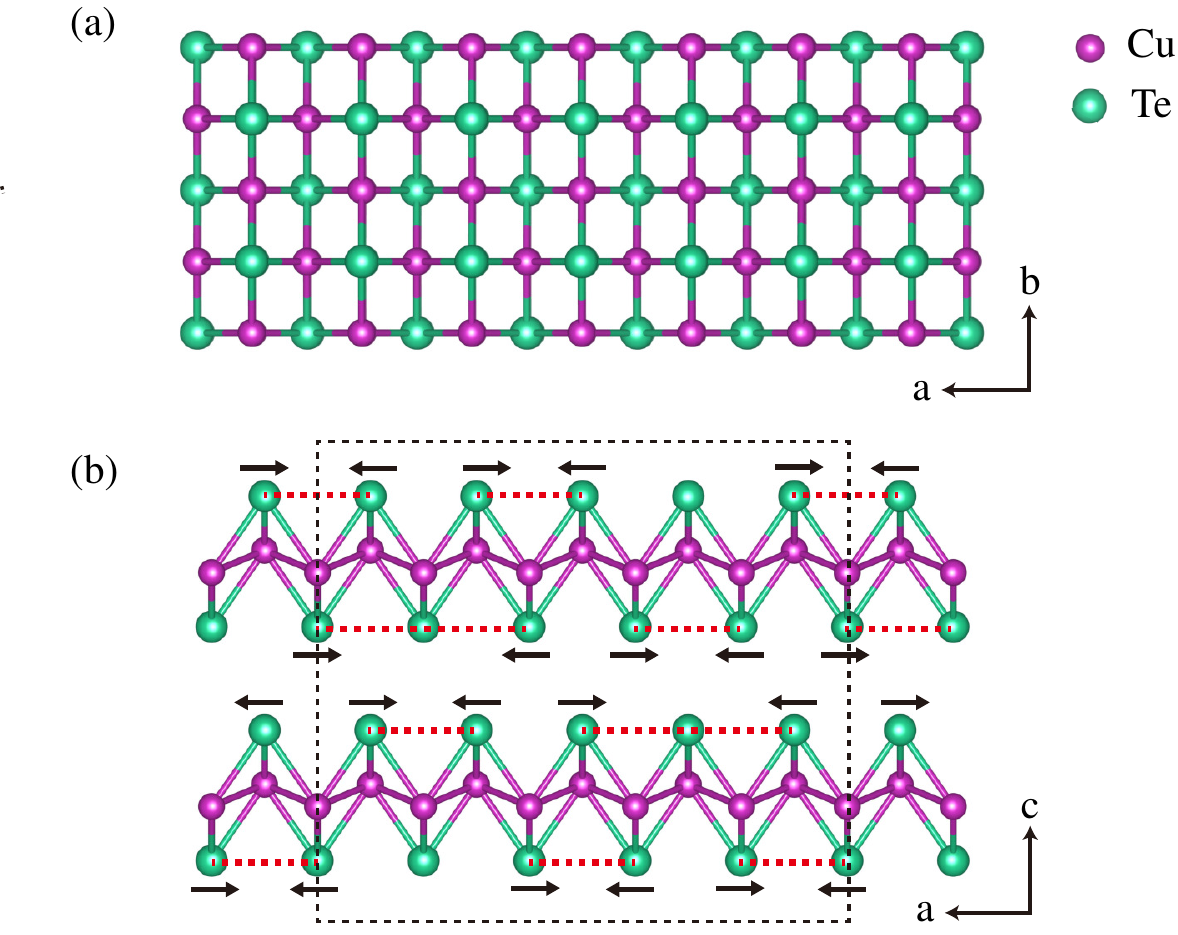}\\[1pt] % insert figure
		\caption{(Color online) Crystal structures of CuTe. (a) Top view and (b) side view of the non-CDW phase. Arrows indicate the movement of Te atoms in the CDW phase.}
		\label{fig:1}
	\end{center}
\end{figure}

CuTe is composed of a stack of edge-sharing CuTe$_{4}$-tetrahedra layer by layer with space group Pmmn and orthorhombic lattice constants of $a=3.124$ $\mathrm{\AA}$, $b=4.086$ $\mathrm{\AA}$, $c=6.946$ $\mathrm{\AA}$~\cite{seong1994te}.  Figure.~\ref{fig:1}(a) shows the top view of the crystal of CuTe projected on the ab-plane, where Cu atoms and Te atoms form a rectangle shape lattice. In the normal state, the Te atoms form a quasi-1D chain along the a-axis where the bond length is shorter than that along the b-axis, which runs above and below the puckered Cu layers with an equal Te-Te distance as shown in Fig.~\ref{fig:1}(b). Below transition temperature, units of two or three Te atoms (marked by red dashed line) with shorter distances (movement of atoms indicated by black arrows in Fig.~\ref{fig:1}(b)) alternate with single Te atom along a-axis.  Meanwhile,  the modulation of the Te position forces corrugated Cu layers to have a slight distortion along the c-axis, resulting in a 5$\times$1$\times$2 superstructure~\cite{stolze2013cute}.

\begin{figure}[htbp]
	\begin{center}
		\includegraphics[clip, width=0.35\textwidth]{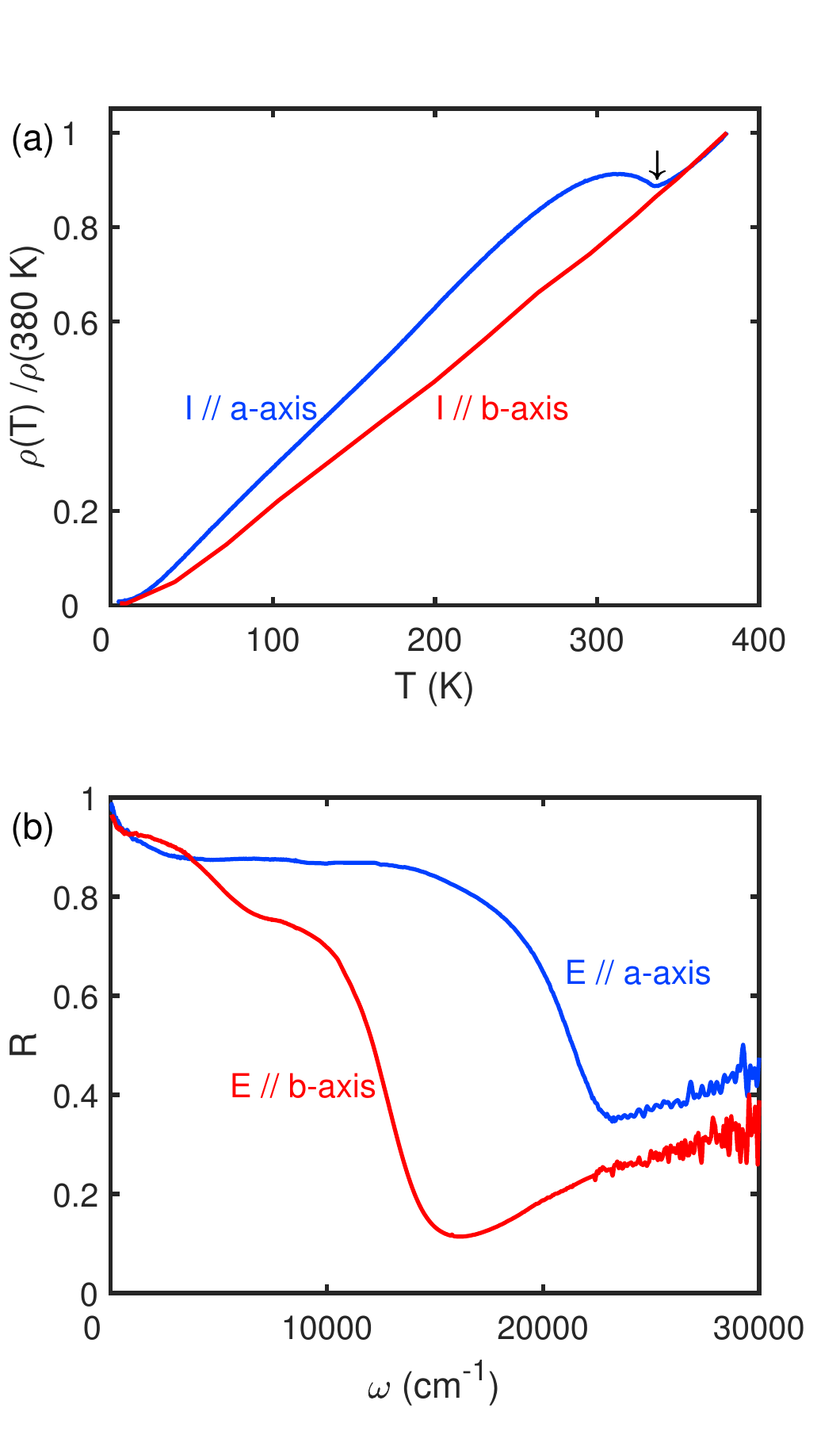}\\[1pt] % insert figure
		\caption{(Color online) (a) The temperature-dependent resistivity of CuTe measured with currents flowing along the crystallographic a- and b-axes, normalized at 380K. (b) Optical reflectivity R($\omega$) in a broad range at 345K, along the a- and b-axes, respectively.}
		\label{fig:2}
	\end{center}
\end{figure}

\begin{figure*}[htpb]
	\begin{center}
		\includegraphics[clip, width=0.8\textwidth]{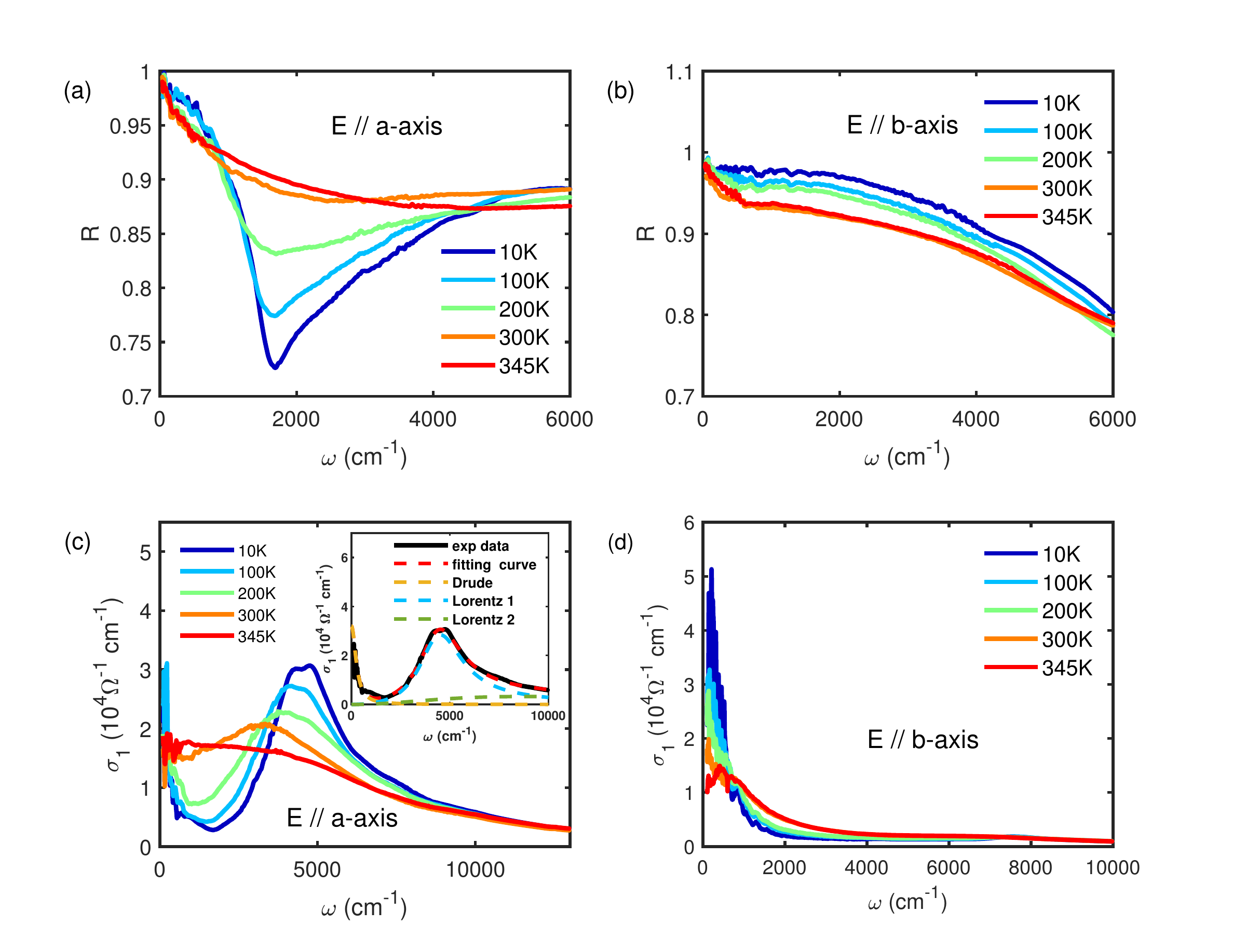}\\[1pt] % insert figure
		\caption{(Color online) The temperature-dependent optical reflectivity R($\omega$) along the a-(a) and b-axes (b), respectively. The optical conductivity $\sigma_1$ at different temperatures along the a-(c) and b-axes (d), respectively. The inset of (c) displays $\sigma_1 (\omega)$ along the a-axis at 10 K together with a Drude-Lorentz fit.}
		\label{fig:3}
	\end{center}
\end{figure*}
	
The temperature-dependent resistivity measurement was performed with a standard four-probe method using a Quantum Design PPMS. Figure~\ref{fig:2}(a) displays the resistivity with currents flowing along the crystallographic a- and b-axes, respectively.  When the electric current flows along the a-axis, the resistivity shows metallic behavior in the entire measured range except a sudden upturn appears at 335 K. This resistivity anomaly is a result of CDW transition, which is consistent with the previous report~\cite{zhang2018evidence}. In contrast,  we didn't find a resistivity anomaly along the b-axis within our measurement resolution. The resistivity data suggest that the CDW order is completely absent in the b-axis, which is quite rare in a highly 2D system.

We performed polarized optical reflectance measurement with the $\textbf{E} \parallel$ the a-axis and the $\textbf{E} \parallel$ b-axis, respectively, on a Bruke 80V Fourier transform infrared spectrometer in the frequency range from 30 to 30000 cm$^{-1}$. An in-suit gold and aluminum overcoating technique was used to get the reflectivity R($\omega$). At 345 K, the reflectivity along the a-axis and b-axis over a broad frequency is plotted in  Fig.~\ref{fig:2}(b).
  Both directions show good metallic behavior in agreement with the dc-resistivity measurement, since reflectivity reaches almost unity at low frequency with a large plasma edge.
Note that, the reflectivity shows a tremendous difference in edge frequency, suggesting that this compound is highly anisotropic.  The edge frequency, being referred to usually as the "screened" plasma frequency, is related to the density $n$ and effective mass $m^{*}$ of free carriers by $\omega_p^{'2}\propto n/m^{*}$. The larger plasma frequency along the a-axis is consistent with the previous report that the conductivity along the chain direction is larger than conductivity perpendicular to the chain at room temperature~\cite{gruner1988dynamics}.

The temperature-dependent reflectivity along the a- and b-axes is displayed in Fig.~\ref{fig:3}(a) and ~\ref{fig:3}(b), respectively.
Let us first discuss optical response along the a-axis.
Firstly, the low-energy reflectivity reaches almost unity. Additionally, when temperature decreases, the low-energy reflectivity increases. Both features show that the sample has a highly metallic nature, in agreement with the dc-resistivity measurement. When the temperature decreases, the most significant feature in spectra $R(\omega)$ is the substantial suppression in the mid-infrared region, which is a typical optical signature of a charge gap formation corresponding to CDW phase transition. The reflectivity exhibits a steep plasma edge around 1800 cm$^{-1}$, revealing that the density of free carriers is substantially suppressed. In the meantime, the scattering rate is also significantly reduced. The dip becomes more dramatic at low temperatures, suggesting that the energy gap is further enhanced.
\begin{figure*}[htbp]
	\begin{center}
		\includegraphics[clip, width=1\textwidth]{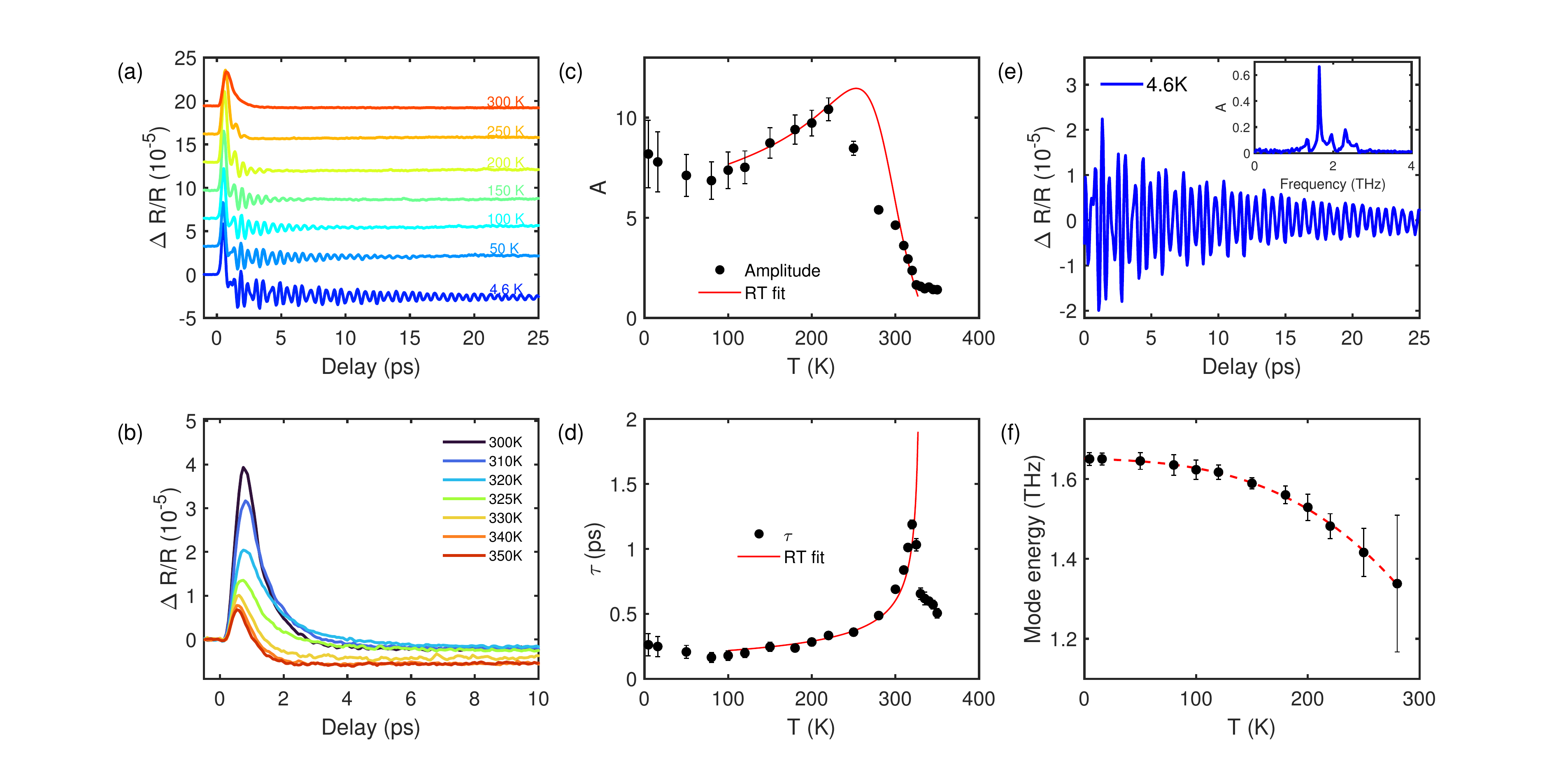}\\[1.2pt] % insert figure
		\caption{(Color online) (a) and (b) The photoinduced reflectivity $\Delta R/R$ along the a-axis as a function of time delay at several selected temperatures. Lines in panel (a) are vertically shifted for clarity. (c) and (d) The amplitude A and relaxation time $\tau$ of the fast decay process extracted from the exponential fitting to the experimental data of $\Delta R /R$ at different temperatures. Error bars represent the standard deviation of the fit. The red curves are fitting curves from Eq. \ref{Eq:2} and \ref{Eq:3} below the phase transition temperatures.  (e) The oscillation part of $\Delta R/R$ at 4.6 K after subtracting the decay background. The inset presents the fast Fourier transformation of the oscillation. (f) The temperature-dependent frequencies of the CDW amplitude mode. The red dashed line is a guide to the eye.}
		\label{fig:4}
	\end{center}
\end{figure*}

The real part of conductivity was derived from R($\omega$) through Kramers-Kronig transformation, as shown in Fig.~\ref{fig:3}(c). The conductivity consists of a Drude peak at zero-frequency and several other Lorentz peaks at a finite frequency in the whole measurement temperature range. At high temperatures, the Drude peak is rather broad demonstrating a large scattering rate of the free carriers. Upon entering the low-temperature phase, the spectral weight of the Drude peak is substantially removed and transfers to higher energies to form a broad peak around 4000 cm$^{-1}$. The Drude peak narrows continuously with decreasing T, implying suppressed quasiparticle scattering. Meanwhile, the positions of the Lorentz peak move to a higher frequency as a result of the further enhancement of the energy gap. In fact, these behaviors are supposed to be the contribution of the case I coherent factor for density wave order~\cite{dressel2002electrodynamics}. We can roughly estimate the scale of the CDW gap through the edge position of the peak, which locates at 4000 cm$^{-1}$(500 meV). In an earlier ARPES measurement, an anisotropic energy gap with a maximum value of 190 meV was identified. It should be noted that ARPES measures the gap relative to the Fermi level, while optical spectroscopy measures the interband transition from the occupied state to the unoccupied state above the Fermi level. The gap value determined by optics usually doubles the energy gap obtained by ARPES if the Fermi level is in the center of the gap. Nevertheless, the energy gap determined by optics can only be taken as a rough estimation because it is hard to determine accurately the conductivity edge. Taking the transitions temperature $T_c$=335 K, we can get the ratio $2\Delta/k_BT_c=17$, which is similar to that reported in other CDW materials, e.g., TbTe$_3$~\cite{PhysRevB.89.075114},  and is much larger than the weak coupling theory prediction~\cite{gruner2018density}.

To get more information quantitatively, we use the Drude-Lorentz model to decompose the conductivity into different components:

\begin{align}\label{Eq:1}
\sigma_1(\omega)=\frac{\omega_p^2}{4\pi} \frac{\Gamma_D}{\omega^2+\Gamma_D^2}
+\sum_j \frac{S_j^2}{4\pi} \frac{\Gamma_j\omega^2}{(\omega_j^2-\omega^2)^2+\omega^2\Gamma_j^2}
\end{align}

\noindent The first term is Drude components attributed to the intraband transition of itinerant carriers, while the second term is Lorentz components correspond to the interband transition and excitation across energy gaps~\cite{PhysRevB.101.205112}. The inset of Fig.~\ref{fig:3}(c) presents the fitting results along a-axis at 10 K.  The spectrum of 345 K can be well reproduced by one Drude and one Lorentz term. At low temperature, one additional Lorentz term needs to be added to reproduce the peak at 4500 cm$^{-1}$. As mentioned above , the variation of plasma frequency $\omega_p$ is proportional to $\sqrt{n/m^*}$. From our fitting date, the plasma frequency varies roughly from 48000 cm$^{-1}$ at 345 K to 22000 cm$^{-1}$ at 10 K. Assuming that the effective mass of free carriers is constant at different temperatures, the reduction of the plasma frequency indicates that 78$\% $ free carriers are removed by the opening of CDW energy gap. Meanwhile, the scattering rate $\gamma$, i.e. the width of the Drude peak, decreases dramatically from 2100 cm$^{-1}$ to 265 cm $^{-1}$ at low temperature.   Consequently, it is reasonable that the compound still behaves like metal even though more than half of free carriers are lost due to the formation of a CDW energy gap.

On the contrary, the reflectivity and the real part of optical conductivity along the b-axis, as shown in Fig.~\ref{fig:3}(b) and (d), do not exhibit these behaviors, but just behave like a simple metal at all temperatures investigated. The scattering rate decreases monotonously upon cooling. No signatures of gap formation along the b-axis in optical spectroscopy, consisting with the CDW formation along the chain direction.

In order to get further information on the CDW transition, we performed ultrafast pump-probe measurement on this compound, which has been proven to be particularly powerful in detecting both the single-particle excitations across small energy gaps~\cite{demsar1999superconducting, chia2010ultrafast} and the collective excitations relevant to long-range ordering~\cite{yusupov2008single, albrecht1992time, qi2013measurement}. We used a Ti: sapphire oscillator as the light source for both pump and probe beams, which can produce 800 nm laser pulses with 100 fs width and 80 MHz repetition rate. The fluence of the pump beam is about 4 $\mu$J/cm$^2$, while the fluence of the probe beam is ten times lower. The polarization direction of the pump and probe pulse are perpendicular to each other, in order to eliminate the interference. We measured photoinduced transient reflectivity $\Delta R/R$ with the electric field $\textbf{E}$ direction of the probe beam parallel to the crystallographic a-axis.

The time-dependent $\Delta R/R$ along the a-axis at several selected temperatures are displayed in Fig. ~\ref{fig:4}(a) and ~\ref{fig:4}(b). Each spectrum initially increases in a very short time induced by the pump excitation, then drops back to the equilibrium state within picoseconds. The relaxation process can be described by a single exponential decay and a constant $\Delta R/R=A \mathrm{exp}(-t/\tau)+C$, where $A$ represents the amplitude of the photoinduced reflectivity change and $\tau$ stands for the relaxation time for photon excited carriers decaying to their original states.  The constant C indicates a long relaxation time arising from thermal diffusion,  where energy from the electrons transforms to the environment.

Above the transition temperature, the amplitude A and relaxation time $\tau$ depend mildly on temperature, while both of them are strongly enhanced by entering the CDW state. The temperature-dependent $A$ and $\tau$ can be extracted by single exponential fitting for various temperatures, which is plotted in Fig.~\ref{fig:4}(c) and \ref{fig:4}(d), respectively. The amplitude and relaxation time display anomalies near the transition temperature. We also tried to fit the $\Delta R/R$ using a convolution between an exponential decay function and a Gaussian response function and found that the fitting results were almost unchanged. The amplitude A drops sharply, while the relaxation time $\tau$ shows a continuous divergence at 330 K. This anomaly behavior of the relaxation process can arise from the energy gap opening, as will be explained in detail later.

Here, we employ the Rothwarf-Taylor (RT) model to interpret the decay of the photoexcited quasiparticles (QPs)~\cite{rothwarf1967measurement}. The RT model was first proposed to resolve the ultrafast relaxation mechanism in superconductivity. It was proved to be applicable for a wide range of materials with a gap opening in the density of states. Photoexcitation would generate a large number of QPs, which would decay via electron-electron or electron-phonon interactions towards to initial equilibrium state.  The formation of an energy gap would introduce a bottleneck effect, which significantly impedes the relaxation of QPs. The bottleneck effect is that, when QPs recombine across an energy gap, high-energy phonons would be generated, which could also induce a large number QPs in return, leading to a long relaxation time. Based on this model, the density of thermally activated quasiparticles $n_T$ can be obtained via the amplitude $A(T)$, $n_T \propto [A(T)/A(T\rightarrow0)]^{-1}-1$. Considering the relationship between $n_T$ and $\Delta(T)$,
$n_T \propto \sqrt{\Delta(T)T}exp[-\Delta(T)/T]$, where $ \Delta(T)=\Delta(0)\sqrt{1-T/T_c}$, we can get~\cite{PhysRevB.59.1497}

\begin{align}\label{Eq:2}
A(T)\propto \frac{\Phi/(\Delta(T)+k_BT/2)}
{1+\gamma \sqrt{2k_BT/\Delta(T)}exp[-\Delta(T)/k_BT]}
\end{align}

\noindent where $\Phi$ is the pump fluence and $\gamma$ is a fitting parameter.
\par As temperature rises, more and more low-energy phonons become available to break condensed carriers with a reduction of the energy gap, which causes the relaxation time of QPs to diverge at the transition temperature.  In the RT model, the relaxation rate near $T_c$ is dominated by the energy transfer from high frequency phonons with $\omega > 2 \Delta$ to low frequency phonons with $\omega < 2 \Delta$. The phonon relaxation rate can be expressed as~\cite{PhysRevB.59.1497}

\begin{align}\label{Eq:3}
\frac{1}{\tau}= \frac{12\Gamma_{\omega}k_BT^{'}\Delta(T)}
{\hbar \omega^2}
\end{align}

\noindent where $\Gamma_{\omega}$ is Raman phonon linewidth and $T^{'}$ is the QPs temperature. 
Apparently,  $\tau(T) \propto \Delta(T)^{-1}$ in the vicinity of $T_c$; thus a divergence is expected at the phase transition temperature where the energy gap starts to open. Qualitatively, these equations can reproduce the measurement results around $T_c$. In the high temperature limit, we take $\omega = 1.65$ THz as the phonons frequency; $T^{\prime}=330$ K as the QP temperature. The best fits yield the phenomenological fitting parameter $\gamma \approx 100$ and $\Gamma_\omega \approx 10$ $cm^{-1}$, as shown in Fig.~\ref{fig:4}(c), (d).

On the other hand, the transient reflectivity $\Delta R/R$ exhibits pronounced oscillatory signals at low temperatures arising from coherent phonon. The lifetime and periods of coherent phonon change with temperature variation. To analyze the component quantitatively, we first subtract the exponential fitting part, then take a fast Fourier transformation of the residual oscillation part. As an example, Fig.~\ref{fig:4}(e) displays $\Delta R/R$ data at 4.6 K as a function of time delay within 25 ps after subtracting the decay background. The inset shows the Fourier transform of the residual part.  Five distinct peaks are observed in the Fourier spectrum at 1.32, 1.65, 1.96, 2.31, 2.59 THz. Based on the previous reports, the strength of CDW amplitude mode is much large than other phonon modes, and its frequency would exhibit softening on approaching $T_c$. Indeed, here the mode at 1.65 THz has a much larger strength and shows significant softening at elevated temperature as shown in Fig.~\ref{fig:4}(f). Compared with the 1.65-THz mode, the other 4 modes decay rapidly with the increase of temperature. We can no longer resolve them above 150 K, while their energy keeps nearly unchanged within our measurement resolution. Therefore, we identify the 1.65-THz phonon as a CDW amplitude mode.  In principle, the 1.65-THz phonon should be softened to zero at the phase transition temperature. However, since the phonon mode is heavily damped at elevated temperatures as reported before, the CDW amplitude mode only softens to 1.34 THz, and it could not be resolved above 280 K.

\section{\label{sec:level4}CONCLUSION}

To conclude, by combining transport, optical spectroscopy, and ultrafast pump-probe spectroscopy, we provide further evidence of anisotropic quasi-1D CDW phase transition in CuTe single crystal.  A strong anomaly near 335 K in the resistivity is seen only along the a-axis, while the feature is completely absent in the b-axis. Our polarized optical measurement shows that the optical responses along a- and b-axes are very different, where the plasma frequency with $\textbf{E} \parallel$ a-axis is much higher than that $\textbf{E} \parallel$ b-axis. When $\textbf{E} \parallel$ a-axis, the real part of optical conductivity clearly reveals the formation of an energy gap with associated spectral weight change. On the contrary, the optical response behaves like a simple metal at all measured temperatures for  $\textbf{E} \parallel$ b-axis.  Furthermore, the amplitude and relaxation time extracted from the photoinduced reflectivity shows a sudden enhancement, consistent well with the RT model prediction, which also confirms the opening of a single-particle energy gap. Meanwhile, we identify the CDW amplitude mode related to the collective excitation of CDW. The spectroscopic information revealed by our anisotropic optical measurement should be very helpful for understanding the electronic properties and related physics. Taking account of the rather high CDW phase transition temperature (above room temperature), CuTe offered an ideal system to study the CDW physics as well as its interplay with superconductivity.

\section*{ACKNOWLEDGMENTS}
This work was supported by the National Key Research and Development Program of China (No. 2017YFA0302904) and National Natural Science Foundation of China (No. 11888101).
\bibliography{Ref}

\begin{thebibliography}{29}%
\makeatletter
\providecommand \@ifxundefined [1]{%
 \@ifx{#1\undefined}
}%
\providecommand \@ifnum [1]{%
 \ifnum #1\expandafter \@firstoftwo
 \else \expandafter \@secondoftwo
 \fi
}%
\providecommand \@ifx [1]{%
 \ifx #1\expandafter \@firstoftwo
 \else \expandafter \@secondoftwo
 \fi
}%
\providecommand \natexlab [1]{#1}%
\providecommand \enquote  [1]{``#1''}%
\providecommand \bibnamefont  [1]{#1}%
\providecommand \bibfnamefont [1]{#1}%
\providecommand \citenamefont [1]{#1}%
\providecommand \href@noop [0]{\@secondoftwo}%
\providecommand \href [0]{\begingroup \@sanitize@url \@href}%
\providecommand \@href[1]{\@@startlink{#1}\@@href}%
\providecommand \@@href[1]{\endgroup#1\@@endlink}%
\providecommand \@sanitize@url [0]{\catcode `\\12\catcode `\$12\catcode
  `\&12\catcode `\#12\catcode `\^12\catcode `\_12\catcode `\%12\relax}%
\providecommand \@@startlink[1]{}%
\providecommand \@@endlink[0]{}%
\providecommand \url  [0]{\begingroup\@sanitize@url \@url }%
\providecommand \@url [1]{\endgroup\@href {#1}{\urlprefix }}%
\providecommand \urlprefix  [0]{URL }%
\providecommand \Eprint [0]{\href }%
\providecommand \doibase [0]{http://dx.doi.org/}%
\providecommand \selectlanguage [0]{\@gobble}%
\providecommand \bibinfo  [0]{\@secondoftwo}%
\providecommand \bibfield  [0]{\@secondoftwo}%
\providecommand \translation [1]{[#1]}%
\providecommand \BibitemOpen [0]{}%
\providecommand \bibitemStop [0]{}%
\providecommand \bibitemNoStop [0]{.\EOS\space}%
\providecommand \EOS [0]{\spacefactor3000\relax}%
\providecommand \BibitemShut  [1]{\csname bibitem#1\endcsname}%
\let\auto@bib@innerbib\@empty
%</preamble>
\bibitem [{\citenamefont {Peierls}\ and\ \citenamefont
  {Peierls}(1955)}]{peierls1955quantum}%
  \BibitemOpen
  \bibfield  {author} {\bibinfo {author} {\bibfnamefont {R.~E.}\ \bibnamefont
  {Peierls}}\ and\ \bibinfo {author} {\bibfnamefont {R.~S.}\ \bibnamefont
  {Peierls}},\ }\href@noop {} {\emph {\bibinfo {title} {Quantum theory of
  solids}}}\ (\bibinfo  {publisher} {Oxford University Press},\ \bibinfo {year}
  {1955})\BibitemShut {NoStop}%
\bibitem [{\citenamefont {Gruner}(2018)}]{gruner2018density}%
  \BibitemOpen
  \bibfield  {author} {\bibinfo {author} {\bibfnamefont {G.}~\bibnamefont
  {Gruner}},\ }\href@noop {} {\emph {\bibinfo {title} {Density waves in
  solids}}}\ (\bibinfo  {publisher} {CRC press},\ \bibinfo {year}
  {2018})\BibitemShut {NoStop}%
\bibitem [{\citenamefont {Ando}\ \emph {et~al.}(2005)\citenamefont {Ando},
  \citenamefont {Yokoya}, \citenamefont {Ishizaka}, \citenamefont {Tsuda},
  \citenamefont {Kiss}, \citenamefont {Shin}, \citenamefont {Eguchi},
  \citenamefont {Nohara},\ and\ \citenamefont {Takagi}}]{Ando_2005}%
  \BibitemOpen
  \bibfield  {author} {\bibinfo {author} {\bibfnamefont {H.}~\bibnamefont
  {Ando}}, \bibinfo {author} {\bibfnamefont {T.}~\bibnamefont {Yokoya}},
  \bibinfo {author} {\bibfnamefont {K.}~\bibnamefont {Ishizaka}}, \bibinfo
  {author} {\bibfnamefont {S.}~\bibnamefont {Tsuda}}, \bibinfo {author}
  {\bibfnamefont {T.}~\bibnamefont {Kiss}}, \bibinfo {author} {\bibfnamefont
  {S.}~\bibnamefont {Shin}}, \bibinfo {author} {\bibfnamefont {T.}~\bibnamefont
  {Eguchi}}, \bibinfo {author} {\bibfnamefont {M.}~\bibnamefont {Nohara}}, \
  and\ \bibinfo {author} {\bibfnamefont {H.}~\bibnamefont {Takagi}},\ }\href
  {\doibase 10.1088/0953-8984/17/32/007} {\bibfield  {journal} {\bibinfo
  {journal} {Journal of Physics: Condensed Matter}\ }\textbf {\bibinfo {volume}
  {17}},\ \bibinfo {pages} {4935} (\bibinfo {year} {2005})}\BibitemShut
  {NoStop}%
\bibitem [{\citenamefont {Johannes}\ and\ \citenamefont
  {Mazin}(2008)}]{PhysRevB.77.165135}%
  \BibitemOpen
  \bibfield  {author} {\bibinfo {author} {\bibfnamefont {M.~D.}\ \bibnamefont
  {Johannes}}\ and\ \bibinfo {author} {\bibfnamefont {I.~I.}\ \bibnamefont
  {Mazin}},\ }\href {\doibase 10.1103/PhysRevB.77.165135} {\bibfield  {journal}
  {\bibinfo  {journal} {Phys. Rev. B}\ }\textbf {\bibinfo {volume} {77}},\
  \bibinfo {pages} {165135} (\bibinfo {year} {2008})}\BibitemShut {NoStop}%
\bibitem [{\citenamefont {Laverock}\ \emph {et~al.}(2005)\citenamefont
  {Laverock}, \citenamefont {Dugdale}, \citenamefont {Major}, \citenamefont
  {Alam}, \citenamefont {Ru}, \citenamefont {Fisher}, \citenamefont {Santi},\
  and\ \citenamefont {Bruno}}]{PhysRevB.71.085114}%
  \BibitemOpen
  \bibfield  {author} {\bibinfo {author} {\bibfnamefont {J.}~\bibnamefont
  {Laverock}}, \bibinfo {author} {\bibfnamefont {S.~B.}\ \bibnamefont
  {Dugdale}}, \bibinfo {author} {\bibfnamefont {Z.}~\bibnamefont {Major}},
  \bibinfo {author} {\bibfnamefont {M.~A.}\ \bibnamefont {Alam}}, \bibinfo
  {author} {\bibfnamefont {N.}~\bibnamefont {Ru}}, \bibinfo {author}
  {\bibfnamefont {I.~R.}\ \bibnamefont {Fisher}}, \bibinfo {author}
  {\bibfnamefont {G.}~\bibnamefont {Santi}}, \ and\ \bibinfo {author}
  {\bibfnamefont {E.}~\bibnamefont {Bruno}},\ }\href {\doibase
  10.1103/PhysRevB.71.085114} {\bibfield  {journal} {\bibinfo  {journal} {Phys.
  Rev. B}\ }\textbf {\bibinfo {volume} {71}},\ \bibinfo {pages} {085114}
  (\bibinfo {year} {2005})}\BibitemShut {NoStop}%
\bibitem [{\citenamefont {Scott}(1974)}]{RevModPhys.46.83}%
  \BibitemOpen
  \bibfield  {author} {\bibinfo {author} {\bibfnamefont {J.~F.}\ \bibnamefont
  {Scott}},\ }\href {\doibase 10.1103/RevModPhys.46.83} {\bibfield  {journal}
  {\bibinfo  {journal} {Rev. Mod. Phys.}\ }\textbf {\bibinfo {volume} {46}},\
  \bibinfo {pages} {83} (\bibinfo {year} {1974})}\BibitemShut {NoStop}%
\bibitem [{\citenamefont {Du}\ \emph {et~al.}(2014)\citenamefont {Du},
  \citenamefont {Yuan}, \citenamefont {Duan}, \citenamefont {Wang},
  \citenamefont {Hu},\ and\ \citenamefont {Li}}]{PhysRevB.90.104414}%
  \BibitemOpen
  \bibfield  {author} {\bibinfo {author} {\bibfnamefont {X.}~\bibnamefont
  {Du}}, \bibinfo {author} {\bibfnamefont {R.}~\bibnamefont {Yuan}}, \bibinfo
  {author} {\bibfnamefont {L.}~\bibnamefont {Duan}}, \bibinfo {author}
  {\bibfnamefont {C.}~\bibnamefont {Wang}}, \bibinfo {author} {\bibfnamefont
  {Y.}~\bibnamefont {Hu}}, \ and\ \bibinfo {author} {\bibfnamefont
  {Y.}~\bibnamefont {Li}},\ }\href {\doibase 10.1103/PhysRevB.90.104414}
  {\bibfield  {journal} {\bibinfo  {journal} {Phys. Rev. B}\ }\textbf {\bibinfo
  {volume} {90}},\ \bibinfo {pages} {104414} (\bibinfo {year}
  {2014})}\BibitemShut {NoStop}%
\bibitem [{\citenamefont {Chen}\ \emph {et~al.}(2017)\citenamefont {Chen},
  \citenamefont {Zhang}, \citenamefont {Zhang}, \citenamefont {Dong},\ and\
  \citenamefont {Wang}}]{PhysRevLett.118.107402}%
  \BibitemOpen
  \bibfield  {author} {\bibinfo {author} {\bibfnamefont {R.~Y.}\ \bibnamefont
  {Chen}}, \bibinfo {author} {\bibfnamefont {S.~J.}\ \bibnamefont {Zhang}},
  \bibinfo {author} {\bibfnamefont {M.~Y.}\ \bibnamefont {Zhang}}, \bibinfo
  {author} {\bibfnamefont {T.}~\bibnamefont {Dong}}, \ and\ \bibinfo {author}
  {\bibfnamefont {N.~L.}\ \bibnamefont {Wang}},\ }\href {\doibase
  10.1103/PhysRevLett.118.107402} {\bibfield  {journal} {\bibinfo  {journal}
  {Phys. Rev. Lett.}\ }\textbf {\bibinfo {volume} {118}},\ \bibinfo {pages}
  {107402} (\bibinfo {year} {2017})}\BibitemShut {NoStop}%
\bibitem [{\citenamefont {Lin}\ \emph {et~al.}(2020)\citenamefont {Lin},
  \citenamefont {Shi}, \citenamefont {Wang}, \citenamefont {Zhang},
  \citenamefont {Liu}, \citenamefont {Hu}, \citenamefont {Dong}, \citenamefont
  {Wu},\ and\ \citenamefont {Wang}}]{PhysRevB.101.205112}%
  \BibitemOpen
  \bibfield  {author} {\bibinfo {author} {\bibfnamefont {T.}~\bibnamefont
  {Lin}}, \bibinfo {author} {\bibfnamefont {L.~Y.}\ \bibnamefont {Shi}},
  \bibinfo {author} {\bibfnamefont {Z.~X.}\ \bibnamefont {Wang}}, \bibinfo
  {author} {\bibfnamefont {S.~J.}\ \bibnamefont {Zhang}}, \bibinfo {author}
  {\bibfnamefont {Q.~M.}\ \bibnamefont {Liu}}, \bibinfo {author} {\bibfnamefont
  {T.~C.}\ \bibnamefont {Hu}}, \bibinfo {author} {\bibfnamefont
  {T.}~\bibnamefont {Dong}}, \bibinfo {author} {\bibfnamefont {D.}~\bibnamefont
  {Wu}}, \ and\ \bibinfo {author} {\bibfnamefont {N.~L.}\ \bibnamefont
  {Wang}},\ }\href {\doibase 10.1103/PhysRevB.101.205112} {\bibfield  {journal}
  {\bibinfo  {journal} {Phys. Rev. B}\ }\textbf {\bibinfo {volume} {101}},\
  \bibinfo {pages} {205112} (\bibinfo {year} {2020})}\BibitemShut {NoStop}%
\bibitem [{\citenamefont {Tsang}\ \emph {et~al.}(1976)\citenamefont {Tsang},
  \citenamefont {Smith},\ and\ \citenamefont {Shafer}}]{PhysRevLett.37.1407}%
  \BibitemOpen
  \bibfield  {author} {\bibinfo {author} {\bibfnamefont {J.~C.}\ \bibnamefont
  {Tsang}}, \bibinfo {author} {\bibfnamefont {J.~E.}\ \bibnamefont {Smith}}, \
  and\ \bibinfo {author} {\bibfnamefont {M.~W.}\ \bibnamefont {Shafer}},\
  }\href {\doibase 10.1103/PhysRevLett.37.1407} {\bibfield  {journal} {\bibinfo
   {journal} {Phys. Rev. Lett.}\ }\textbf {\bibinfo {volume} {37}},\ \bibinfo
  {pages} {1407} (\bibinfo {year} {1976})}\BibitemShut {NoStop}%
\bibitem [{\citenamefont {Hu}\ \emph {et~al.}(2015)\citenamefont {Hu},
  \citenamefont {Zheng}, \citenamefont {Ren}, \citenamefont {Feng},\ and\
  \citenamefont {Li}}]{PhysRevB.91.144502}%
  \BibitemOpen
  \bibfield  {author} {\bibinfo {author} {\bibfnamefont {Y.}~\bibnamefont
  {Hu}}, \bibinfo {author} {\bibfnamefont {F.}~\bibnamefont {Zheng}}, \bibinfo
  {author} {\bibfnamefont {X.}~\bibnamefont {Ren}}, \bibinfo {author}
  {\bibfnamefont {J.}~\bibnamefont {Feng}}, \ and\ \bibinfo {author}
  {\bibfnamefont {Y.}~\bibnamefont {Li}},\ }\href {\doibase
  10.1103/PhysRevB.91.144502} {\bibfield  {journal} {\bibinfo  {journal} {Phys.
  Rev. B}\ }\textbf {\bibinfo {volume} {91}},\ \bibinfo {pages} {144502}
  (\bibinfo {year} {2015})}\BibitemShut {NoStop}%
\bibitem [{\citenamefont {Pouget}\ \emph {et~al.}(1991)\citenamefont {Pouget},
  \citenamefont {Hennion}, \citenamefont {Escribe-Filippini},\ and\
  \citenamefont {Sato}}]{PhysRevB.43.8421}%
  \BibitemOpen
  \bibfield  {author} {\bibinfo {author} {\bibfnamefont {J.~P.}\ \bibnamefont
  {Pouget}}, \bibinfo {author} {\bibfnamefont {B.}~\bibnamefont {Hennion}},
  \bibinfo {author} {\bibfnamefont {C.}~\bibnamefont {Escribe-Filippini}}, \
  and\ \bibinfo {author} {\bibfnamefont {M.}~\bibnamefont {Sato}},\ }\href
  {\doibase 10.1103/PhysRevB.43.8421} {\bibfield  {journal} {\bibinfo
  {journal} {Phys. Rev. B}\ }\textbf {\bibinfo {volume} {43}},\ \bibinfo
  {pages} {8421} (\bibinfo {year} {1991})}\BibitemShut {NoStop}%
\bibitem [{\citenamefont {Smith}\ \emph {et~al.}(1985)\citenamefont {Smith},
  \citenamefont {Kevan},\ and\ \citenamefont {DiSalvo}}]{Smith_1985}%
  \BibitemOpen
  \bibfield  {author} {\bibinfo {author} {\bibfnamefont {N.~V.}\ \bibnamefont
  {Smith}}, \bibinfo {author} {\bibfnamefont {S.~D.}\ \bibnamefont {Kevan}}, \
  and\ \bibinfo {author} {\bibfnamefont {F.~J.}\ \bibnamefont {DiSalvo}},\
  }\href {\doibase 10.1088/0022-3719/18/16/013} {\bibfield  {journal} {\bibinfo
   {journal} {Journal of Physics C: Solid State Physics}\ }\textbf {\bibinfo
  {volume} {18}},\ \bibinfo {pages} {3175} (\bibinfo {year}
  {1985})}\BibitemShut {NoStop}%
\bibitem [{\citenamefont {Moore}\ \emph {et~al.}(2010)\citenamefont {Moore},
  \citenamefont {Brouet}, \citenamefont {He}, \citenamefont {Lu}, \citenamefont
  {Ru}, \citenamefont {Chu}, \citenamefont {Fisher},\ and\ \citenamefont
  {Shen}}]{PhysRevB.81.073102}%
  \BibitemOpen
  \bibfield  {author} {\bibinfo {author} {\bibfnamefont {R.~G.}\ \bibnamefont
  {Moore}}, \bibinfo {author} {\bibfnamefont {V.}~\bibnamefont {Brouet}},
  \bibinfo {author} {\bibfnamefont {R.}~\bibnamefont {He}}, \bibinfo {author}
  {\bibfnamefont {D.~H.}\ \bibnamefont {Lu}}, \bibinfo {author} {\bibfnamefont
  {N.}~\bibnamefont {Ru}}, \bibinfo {author} {\bibfnamefont {J.-H.}\
  \bibnamefont {Chu}}, \bibinfo {author} {\bibfnamefont {I.~R.}\ \bibnamefont
  {Fisher}}, \ and\ \bibinfo {author} {\bibfnamefont {Z.-X.}\ \bibnamefont
  {Shen}},\ }\href {\doibase 10.1103/PhysRevB.81.073102} {\bibfield  {journal}
  {\bibinfo  {journal} {Phys. Rev. B}\ }\textbf {\bibinfo {volume} {81}},\
  \bibinfo {pages} {073102} (\bibinfo {year} {2010})}\BibitemShut {NoStop}%
\bibitem [{\citenamefont {Chen}\ \emph {et~al.}(2014)\citenamefont {Chen},
  \citenamefont {Hu}, \citenamefont {Dong},\ and\ \citenamefont
  {Wang}}]{PhysRevB.89.075114}%
  \BibitemOpen
  \bibfield  {author} {\bibinfo {author} {\bibfnamefont {R.~Y.}\ \bibnamefont
  {Chen}}, \bibinfo {author} {\bibfnamefont {B.~F.}\ \bibnamefont {Hu}},
  \bibinfo {author} {\bibfnamefont {T.}~\bibnamefont {Dong}}, \ and\ \bibinfo
  {author} {\bibfnamefont {N.~L.}\ \bibnamefont {Wang}},\ }\href {\doibase
  10.1103/PhysRevB.89.075114} {\bibfield  {journal} {\bibinfo  {journal} {Phys.
  Rev. B}\ }\textbf {\bibinfo {volume} {89}},\ \bibinfo {pages} {075114}
  (\bibinfo {year} {2014})}\BibitemShut {NoStop}%
\bibitem [{\citenamefont {Stolze}\ \emph {et~al.}(2013)\citenamefont {Stolze},
  \citenamefont {Isaeva}, \citenamefont {Nitsche}, \citenamefont {Burkhardt},
  \citenamefont {Lichte}, \citenamefont {Wolf},\ and\ \citenamefont
  {Doert}}]{stolze2013cute}%
  \BibitemOpen
  \bibfield  {author} {\bibinfo {author} {\bibfnamefont {K.}~\bibnamefont
  {Stolze}}, \bibinfo {author} {\bibfnamefont {A.}~\bibnamefont {Isaeva}},
  \bibinfo {author} {\bibfnamefont {F.}~\bibnamefont {Nitsche}}, \bibinfo
  {author} {\bibfnamefont {U.}~\bibnamefont {Burkhardt}}, \bibinfo {author}
  {\bibfnamefont {H.}~\bibnamefont {Lichte}}, \bibinfo {author} {\bibfnamefont
  {D.}~\bibnamefont {Wolf}}, \ and\ \bibinfo {author} {\bibfnamefont
  {T.}~\bibnamefont {Doert}},\ }\href {\doibase
  https://doi.org/10.1002/anie.201207333} {\bibfield  {journal} {\bibinfo
  {journal} {Angewandte Chemie International Edition}\ }\textbf {\bibinfo
  {volume} {52}},\ \bibinfo {pages} {862} (\bibinfo {year} {2013})}\BibitemShut
  {NoStop}%
\bibitem [{\citenamefont {Zhang}\ \emph {et~al.}(2018)\citenamefont {Zhang},
  \citenamefont {Liu}, \citenamefont {Zhang}, \citenamefont {Deng},
  \citenamefont {Yan}, \citenamefont {Yao}, \citenamefont {Zheng},
  \citenamefont {Schwier}, \citenamefont {Shimada}, \citenamefont {Denlinger},
  \citenamefont {Wu}, \citenamefont {Duan},\ and\ \citenamefont
  {Zhou}}]{zhang2018evidence}%
  \BibitemOpen
  \bibfield  {author} {\bibinfo {author} {\bibfnamefont {K.}~\bibnamefont
  {Zhang}}, \bibinfo {author} {\bibfnamefont {X.}~\bibnamefont {Liu}}, \bibinfo
  {author} {\bibfnamefont {H.}~\bibnamefont {Zhang}}, \bibinfo {author}
  {\bibfnamefont {K.}~\bibnamefont {Deng}}, \bibinfo {author} {\bibfnamefont
  {M.}~\bibnamefont {Yan}}, \bibinfo {author} {\bibfnamefont {W.}~\bibnamefont
  {Yao}}, \bibinfo {author} {\bibfnamefont {M.}~\bibnamefont {Zheng}}, \bibinfo
  {author} {\bibfnamefont {E.~F.}\ \bibnamefont {Schwier}}, \bibinfo {author}
  {\bibfnamefont {K.}~\bibnamefont {Shimada}}, \bibinfo {author} {\bibfnamefont
  {J.~D.}\ \bibnamefont {Denlinger}}, \bibinfo {author} {\bibfnamefont
  {Y.}~\bibnamefont {Wu}}, \bibinfo {author} {\bibfnamefont {W.}~\bibnamefont
  {Duan}}, \ and\ \bibinfo {author} {\bibfnamefont {S.}~\bibnamefont {Zhou}},\
  }\href {\doibase 10.1103/PhysRevLett.121.206402} {\bibfield  {journal}
  {\bibinfo  {journal} {Phys. Rev. Lett.}\ }\textbf {\bibinfo {volume} {121}},\
  \bibinfo {pages} {206402} (\bibinfo {year} {2018})}\BibitemShut {NoStop}%
\bibitem [{\citenamefont {Kuo}\ \emph {et~al.}(2020)\citenamefont {Kuo},
  \citenamefont {Huang}, \citenamefont {Kuo},\ and\ \citenamefont
  {Lue}}]{kuo2020transport}%
  \BibitemOpen
  \bibfield  {author} {\bibinfo {author} {\bibfnamefont {C.~N.}\ \bibnamefont
  {Kuo}}, \bibinfo {author} {\bibfnamefont {R.~Y.}\ \bibnamefont {Huang}},
  \bibinfo {author} {\bibfnamefont {Y.~K.}\ \bibnamefont {Kuo}}, \ and\
  \bibinfo {author} {\bibfnamefont {C.~S.}\ \bibnamefont {Lue}},\ }\href
  {\doibase 10.1103/PhysRevB.102.155137} {\bibfield  {journal} {\bibinfo
  {journal} {Phys. Rev. B}\ }\textbf {\bibinfo {volume} {102}},\ \bibinfo
  {pages} {155137} (\bibinfo {year} {2020})}\BibitemShut {NoStop}%
\bibitem [{\citenamefont {Wang}\ \emph {et~al.}(2021)\citenamefont {Wang},
  \citenamefont {Chen}, \citenamefont {An}, \citenamefont {Zhou}, \citenamefont
  {Zhou}, \citenamefont {Gu}, \citenamefont {Zhang}, \citenamefont {Yang},\
  and\ \citenamefont {Yang}}]{wang2021pressure}%
  \BibitemOpen
  \bibfield  {author} {\bibinfo {author} {\bibfnamefont {S.}~\bibnamefont
  {Wang}}, \bibinfo {author} {\bibfnamefont {X.}~\bibnamefont {Chen}}, \bibinfo
  {author} {\bibfnamefont {C.}~\bibnamefont {An}}, \bibinfo {author}
  {\bibfnamefont {Y.}~\bibnamefont {Zhou}}, \bibinfo {author} {\bibfnamefont
  {Y.}~\bibnamefont {Zhou}}, \bibinfo {author} {\bibfnamefont {C.}~\bibnamefont
  {Gu}}, \bibinfo {author} {\bibfnamefont {L.}~\bibnamefont {Zhang}}, \bibinfo
  {author} {\bibfnamefont {X.}~\bibnamefont {Yang}}, \ and\ \bibinfo {author}
  {\bibfnamefont {Z.}~\bibnamefont {Yang}},\ }\href {\doibase
  10.1103/PhysRevB.103.134518} {\bibfield  {journal} {\bibinfo  {journal}
  {Phys. Rev. B}\ }\textbf {\bibinfo {volume} {103}},\ \bibinfo {pages}
  {134518} (\bibinfo {year} {2021})}\BibitemShut {NoStop}%
\bibitem [{\citenamefont {Seong}\ \emph {et~al.}(1994)\citenamefont {Seong},
  \citenamefont {Albright}, \citenamefont {Zhang},\ and\ \citenamefont
  {Kanatzidis}}]{seong1994te}%
  \BibitemOpen
  \bibfield  {author} {\bibinfo {author} {\bibfnamefont {S.}~\bibnamefont
  {Seong}}, \bibinfo {author} {\bibfnamefont {T.~A.}\ \bibnamefont {Albright}},
  \bibinfo {author} {\bibfnamefont {X.}~\bibnamefont {Zhang}}, \ and\ \bibinfo
  {author} {\bibfnamefont {M.}~\bibnamefont {Kanatzidis}},\ }\href {\doibase
  10.1021/ja00095a036} {\bibfield  {journal} {\bibinfo  {journal} {Journal of
  the American Chemical Society}\ }\textbf {\bibinfo {volume} {116}},\ \bibinfo
  {pages} {7287} (\bibinfo {year} {1994})}\BibitemShut {NoStop}%
\bibitem [{\citenamefont {Gr\"uner}(1988)}]{gruner1988dynamics}%
  \BibitemOpen
  \bibfield  {author} {\bibinfo {author} {\bibfnamefont {G.}~\bibnamefont
  {Gr\"uner}},\ }\href {\doibase 10.1103/RevModPhys.60.1129} {\bibfield
  {journal} {\bibinfo  {journal} {Rev. Mod. Phys.}\ }\textbf {\bibinfo {volume}
  {60}},\ \bibinfo {pages} {1129} (\bibinfo {year} {1988})}\BibitemShut
  {NoStop}%
\bibitem [{\citenamefont {Dressel}\ and\ \citenamefont
  {Grüner}(2002)}]{dressel2002electrodynamics}%
  \BibitemOpen
  \bibfield  {author} {\bibinfo {author} {\bibfnamefont {M.}~\bibnamefont
  {Dressel}}\ and\ \bibinfo {author} {\bibfnamefont {G.}~\bibnamefont
  {Grüner}},\ }\href {\doibase 10.1017/CBO9780511606168} {\emph {\bibinfo
  {title} {Electrodynamics of Solids: Optical Properties of Electrons in
  Matter}}}\ (\bibinfo  {publisher} {Cambridge University Press},\ \bibinfo
  {year} {2002})\BibitemShut {NoStop}%
\bibitem [{\citenamefont {Demsar}\ \emph {et~al.}(1999)\citenamefont {Demsar},
  \citenamefont {Podobnik}, \citenamefont {Kabanov}, \citenamefont {Wolf},\
  and\ \citenamefont {Mihailovic}}]{demsar1999superconducting}%
  \BibitemOpen
  \bibfield  {author} {\bibinfo {author} {\bibfnamefont {J.}~\bibnamefont
  {Demsar}}, \bibinfo {author} {\bibfnamefont {B.}~\bibnamefont {Podobnik}},
  \bibinfo {author} {\bibfnamefont {V.~V.}\ \bibnamefont {Kabanov}}, \bibinfo
  {author} {\bibfnamefont {T.}~\bibnamefont {Wolf}}, \ and\ \bibinfo {author}
  {\bibfnamefont {D.}~\bibnamefont {Mihailovic}},\ }\href {\doibase
  10.1103/PhysRevLett.82.4918} {\bibfield  {journal} {\bibinfo  {journal}
  {Phys. Rev. Lett.}\ }\textbf {\bibinfo {volume} {82}},\ \bibinfo {pages}
  {4918} (\bibinfo {year} {1999})}\BibitemShut {NoStop}%
\bibitem [{\citenamefont {Chia}\ \emph {et~al.}(2010)\citenamefont {Chia},
  \citenamefont {Talbayev}, \citenamefont {Zhu}, \citenamefont {Yuan},
  \citenamefont {Park}, \citenamefont {Thompson}, \citenamefont {Panagopoulos},
  \citenamefont {Chen}, \citenamefont {Luo}, \citenamefont {Wang},\ and\
  \citenamefont {Taylor}}]{chia2010ultrafast}%
  \BibitemOpen
  \bibfield  {author} {\bibinfo {author} {\bibfnamefont {E.~E.~M.}\
  \bibnamefont {Chia}}, \bibinfo {author} {\bibfnamefont {D.}~\bibnamefont
  {Talbayev}}, \bibinfo {author} {\bibfnamefont {J.-X.}\ \bibnamefont {Zhu}},
  \bibinfo {author} {\bibfnamefont {H.~Q.}\ \bibnamefont {Yuan}}, \bibinfo
  {author} {\bibfnamefont {T.}~\bibnamefont {Park}}, \bibinfo {author}
  {\bibfnamefont {J.~D.}\ \bibnamefont {Thompson}}, \bibinfo {author}
  {\bibfnamefont {C.}~\bibnamefont {Panagopoulos}}, \bibinfo {author}
  {\bibfnamefont {G.~F.}\ \bibnamefont {Chen}}, \bibinfo {author}
  {\bibfnamefont {J.~L.}\ \bibnamefont {Luo}}, \bibinfo {author} {\bibfnamefont
  {N.~L.}\ \bibnamefont {Wang}}, \ and\ \bibinfo {author} {\bibfnamefont
  {A.~J.}\ \bibnamefont {Taylor}},\ }\href {\doibase
  10.1103/PhysRevLett.104.027003} {\bibfield  {journal} {\bibinfo  {journal}
  {Phys. Rev. Lett.}\ }\textbf {\bibinfo {volume} {104}},\ \bibinfo {pages}
  {027003} (\bibinfo {year} {2010})}\BibitemShut {NoStop}%
\bibitem [{\citenamefont {Yusupov}\ \emph {et~al.}(2008)\citenamefont
  {Yusupov}, \citenamefont {Mertelj}, \citenamefont {Chu}, \citenamefont
  {Fisher},\ and\ \citenamefont {Mihailovic}}]{yusupov2008single}%
  \BibitemOpen
  \bibfield  {author} {\bibinfo {author} {\bibfnamefont {R.~V.}\ \bibnamefont
  {Yusupov}}, \bibinfo {author} {\bibfnamefont {T.}~\bibnamefont {Mertelj}},
  \bibinfo {author} {\bibfnamefont {J.-H.}\ \bibnamefont {Chu}}, \bibinfo
  {author} {\bibfnamefont {I.~R.}\ \bibnamefont {Fisher}}, \ and\ \bibinfo
  {author} {\bibfnamefont {D.}~\bibnamefont {Mihailovic}},\ }\href {\doibase
  10.1103/PhysRevLett.101.246402} {\bibfield  {journal} {\bibinfo  {journal}
  {Phys. Rev. Lett.}\ }\textbf {\bibinfo {volume} {101}},\ \bibinfo {pages}
  {246402} (\bibinfo {year} {2008})}\BibitemShut {NoStop}%
\bibitem [{\citenamefont {Albrecht}\ \emph {et~al.}(1992)\citenamefont
  {Albrecht}, \citenamefont {Kruse},\ and\ \citenamefont
  {Kurz}}]{albrecht1992time}%
  \BibitemOpen
  \bibfield  {author} {\bibinfo {author} {\bibfnamefont {W.}~\bibnamefont
  {Albrecht}}, \bibinfo {author} {\bibfnamefont {T.}~\bibnamefont {Kruse}}, \
  and\ \bibinfo {author} {\bibfnamefont {H.}~\bibnamefont {Kurz}},\ }\href
  {\doibase 10.1103/PhysRevLett.69.1451} {\bibfield  {journal} {\bibinfo
  {journal} {Phys. Rev. Lett.}\ }\textbf {\bibinfo {volume} {69}},\ \bibinfo
  {pages} {1451} (\bibinfo {year} {1992})}\BibitemShut {NoStop}%
\bibitem [{\citenamefont {Qi}\ \emph {et~al.}(2013)\citenamefont {Qi},
  \citenamefont {Durakiewicz}, \citenamefont {Trugman}, \citenamefont {Zhu},
  \citenamefont {Riseborough}, \citenamefont {Baumbach}, \citenamefont {Bauer},
  \citenamefont {Gofryk}, \citenamefont {Meng}, \citenamefont {Joyce},
  \citenamefont {Taylor},\ and\ \citenamefont
  {Prasankumar}}]{qi2013measurement}%
  \BibitemOpen
  \bibfield  {author} {\bibinfo {author} {\bibfnamefont {J.}~\bibnamefont
  {Qi}}, \bibinfo {author} {\bibfnamefont {T.}~\bibnamefont {Durakiewicz}},
  \bibinfo {author} {\bibfnamefont {S.~A.}\ \bibnamefont {Trugman}}, \bibinfo
  {author} {\bibfnamefont {J.-X.}\ \bibnamefont {Zhu}}, \bibinfo {author}
  {\bibfnamefont {P.~S.}\ \bibnamefont {Riseborough}}, \bibinfo {author}
  {\bibfnamefont {R.}~\bibnamefont {Baumbach}}, \bibinfo {author}
  {\bibfnamefont {E.~D.}\ \bibnamefont {Bauer}}, \bibinfo {author}
  {\bibfnamefont {K.}~\bibnamefont {Gofryk}}, \bibinfo {author} {\bibfnamefont
  {J.-Q.}\ \bibnamefont {Meng}}, \bibinfo {author} {\bibfnamefont {J.~J.}\
  \bibnamefont {Joyce}}, \bibinfo {author} {\bibfnamefont {A.~J.}\ \bibnamefont
  {Taylor}}, \ and\ \bibinfo {author} {\bibfnamefont {R.~P.}\ \bibnamefont
  {Prasankumar}},\ }\href {\doibase 10.1103/PhysRevLett.111.057402} {\bibfield
  {journal} {\bibinfo  {journal} {Phys. Rev. Lett.}\ }\textbf {\bibinfo
  {volume} {111}},\ \bibinfo {pages} {057402} (\bibinfo {year}
  {2013})}\BibitemShut {NoStop}%
\bibitem [{\citenamefont {Rothwarf}\ and\ \citenamefont
  {Taylor}(1967)}]{rothwarf1967measurement}%
  \BibitemOpen
  \bibfield  {author} {\bibinfo {author} {\bibfnamefont {A.}~\bibnamefont
  {Rothwarf}}\ and\ \bibinfo {author} {\bibfnamefont {B.~N.}\ \bibnamefont
  {Taylor}},\ }\href {\doibase 10.1103/PhysRevLett.19.27} {\bibfield  {journal}
  {\bibinfo  {journal} {Phys. Rev. Lett.}\ }\textbf {\bibinfo {volume} {19}},\
  \bibinfo {pages} {27} (\bibinfo {year} {1967})}\BibitemShut {NoStop}%
\bibitem [{\citenamefont {Kabanov}\ \emph {et~al.}(1999)\citenamefont
  {Kabanov}, \citenamefont {Demsar}, \citenamefont {Podobnik},\ and\
  \citenamefont {Mihailovic}}]{PhysRevB.59.1497}%
  \BibitemOpen
  \bibfield  {author} {\bibinfo {author} {\bibfnamefont {V.~V.}\ \bibnamefont
  {Kabanov}}, \bibinfo {author} {\bibfnamefont {J.}~\bibnamefont {Demsar}},
  \bibinfo {author} {\bibfnamefont {B.}~\bibnamefont {Podobnik}}, \ and\
  \bibinfo {author} {\bibfnamefont {D.}~\bibnamefont {Mihailovic}},\ }\href
  {\doibase 10.1103/PhysRevB.59.1497} {\bibfield  {journal} {\bibinfo
  {journal} {Phys. Rev. B}\ }\textbf {\bibinfo {volume} {59}},\ \bibinfo
  {pages} {1497} (\bibinfo {year} {1999})}\BibitemShut {NoStop}%
\end{thebibliography}%

\clearpage

\end{document}